# Long-term Periodicities in the Flux from Low Mass X-ray Binaries

Stefan Dieters[a,1]; Paul O'Neill[b]; Sean Farrell[a] and Ravi Sood[a]

[a]School of PEMS, University of NSW, ADFA, Canberra ACT 2600, Australia
[b]Blackett Laboratory, Imperial College, London SW7 2BW, UK

**Abstract**

Using data from the All Sky Monitor (ASM) on the Rossi X-ray Timing Explorer (RXTE) we have searched for long term periodicities in the X-ray flux of GX 1+4, Sco X-2 (GX 349+2), and GX 339-4. For GX 1+4 we also used data from BATSE and Galactic Centre scans performed by RXTE. We find no evidence for X-ray modulations at the suggested ~304 d orbital period of GX 1+4. However, we find tentative evidence for a periodicity at 420 d to 460 d. An upper limit of 15% peak-to-peak is set on any sinusoidal modulation in the 1.5 – 3.0 keV flux of Sco X-2 for periods in the 30 to 100 d range. For GX 339-4 we confirm the Low State modulation and report the detection of significant low-frequency modulations in both the High State and Very High State. We fail to detect this modulation in the Off State. We show that if the reported orbital period of GX 339-4 lies in the range 0.5 – 1.7 d, then it is not present in the RXTE ASM light curve.

## 1. Introduction

Low mass X-ray binaries (LMXBs) consist of a low mass ($< 2 M_\odot$) more-or-less normal star donating mass via Roche lobe overflow, and an accretion disk to a neutron star or a black hole. For a high magnetic field strength neutron star, X-ray pulsations are observed at the neutron star spin period. All LMXBs show strong, non-periodic variability over all time scales from milliseconds to years. A combination of the timing and spectral variability i.e. *states*, is used to classify LMXBs. The "atoll" and "Z" sources are named after the shape they trace out in an X-ray colour-colour diagram. Systems containing a black hole show yet another set of X-ray temporal and spectral states.

A wide variety of disk fed systems show (quasi-) periodicities longer than the orbital period. Observations of these systems point to such super-orbital periods as being due to changes in viewing geometry as a tilted and/or warped disk precesses because of radiation pressure (Priedhorsky & Holt 1987). The super-orbital modulations that have been detected from LMXBs are generally of lower amplitude and of lower coherence than those from high mass systems such as SMC X-1. Here we present the results of a search for such periodicities in the flux from the Z-source Sco X-2 (GX 349+2), the black hole candidate GX 339-4, and a search for the orbital period of GX 1+4, an X-ray pulsar.

## 2. Observations and General Data Analysis

The Rossi X-ray Timing Explorer (RXTE) carries an All Sky Monitor (ASM) with 3 Scanning Shadow Cameras (SSC). It has been in continuous operation since 1996. Each scan or "dwell" lasts for 90 s. A fit that uses known source positions and a detector response function returns the average flux in three energy bands (A "soft": 1.5 – 3.0, B: "medium" 3.0 – 5.0, and C: "hard" 5.0 – 12.0 keV). This procedure may produce a systematic uncertainty of ~0.1 cts/s in each band. There are slow gain changes in the SSC1 and SSC3 detectors. A discrete gain adjustment to the SSC1 and SSC3 was made on MJD 51548. The SSC2 detector gain is stable.

Our data analysis included measures to avoid or mitigate any trends due to ASM detector gain changes. Also, as we are mostly interested in smooth long-term periodicities we took steps to mitigate the effects of source flaring. This was only necessary for GX 1+4 and Sco X-2. Details specific to the analysis of each source are given in the section for that source. To search for periodicities we used the Lomb-Scargle periodogram (LSP) technique. Power upper limits were set by creating distributions of highest powers over the entire frequency range and finding the 99, 95, 90% levels. The count rates were first scrambled in time, then the periodogram was computed, and the highest power found. This process was repeated 10,000 times. It yields a conservative detection limit as it is really an upper limit to the statistical limit. Where no signal was detected, an upper limit was set by adding different amplitude sinusoids to the data; finding the amplitude that was just above the detection threshold.

## 3. GX 1+4

*3.1 Introduction*

---

[1] Present address: School of Maths and Physics, Univ. of Tasmania, Hobart TAS 7000, Australia.

GX 1+4 is a rare, red-giant / neutron star system. The pulse-period of ~2 min shows the largest period derivative of any X-ray pulsar during long intervals of nearly continuous spin-up or spin-down. Considering the size and evolution of the giant companion, the orbital period must be >100 d and most likely > 260 d. A period of ~304 d has been reported by several authors based on X-ray timing analysis rather than on changes in flux (see Pereira et al. (1999) and references therein). An eccentric orbit as suggested by these timing studies would lead to a flux modulation via variable accretion.

*3.2 Analysis*

Just over 7 years of RXTE/ASM data from MJD 50088 to 52774 were used. Dwells with high $\chi^2$ fit values were removed. The cut-off level for each detector was set at the beginning of the high side tail in the distribution of $\chi^2$ values. This filtering did not have a count rate bias i.e. the removed dwells came from all count rates. A threshold level of $\chi^2 < 1.2$ was used for all 3 detectors respectively. This removed ~7 % of dwells for SSC1 and SSC2 detectors, and ~13 % for the SSC3 detector. The SSC3 detector had a wider distribution of $\chi^2$ values. Next, the gain changes were removed by fitting 20 d averages with separate quadratics before and after gain adjustment for each detector and each energy band. Each trend was subtracted from the single dwell data. This was the standard filtering applied to all the data for this source. In order to specifically search for smooth flux variations additional filtering was undertaken. Before trend removal high count-rate individual dwells were removed. A further 12 % of the data was thus rejected. The combination of this filtering and binning into 1 d or 10 d averages removed any flaring. Further analysis was preformed on standard and peak filtered data separately. Data were binned into 1 d, and 10 d averages for each energy channel, SUM (all energies) and sums of A+B, B+C energy bands. LSP analysis was performed on the 1 d, and 10 d binned light-curves in each energy band. Generally a factor of 20 over-sampling was used so as not to miss any peaks in the periodogram. Upper limits were set, and a percentage limit was set using test sinusoids with a period of 300 d.

*3.3 Results*

Analysis of the flare-filtered 10 d averaged ASM data showed no significant periodicities over the 20 – 2000 d range in any of the individual or summed energy ranges. In particular there was no evidence for a period at ~304 d. A 1σ upper limit of 16% peak-to-peak (p-p) amplitude was set by phase folding the 1.5 – 12.0 keV data at this period. When flares were included, possible (95% confidence) periods were detected at 440 d and 788 d in the SUM, B+C, and C channels. The p-p amplitudes, (1.5 – 12.0 keV), were ~25% and ~19% for the 440 d and 788 d periods respectively. Neither of these periods appeared in the window function for the ASM data. Fig. 1 shows the periodogram of the 1 d averaged data with standard filtering (i.e. no flares removed) over the full energy range.

A close examination of the ASM light curve shows several instances where GX 1+4 apparently underwent short (≤ 10 d) flares. These flares may be periodic as with transient Be / X-ray pulsar systems with eccentric orbits. A period search based on the times of possible flares was suggested because the periodogram of the data that included flares showed more significant peaks. Times of high (> 10 cts/s) count rate dwells were searched for periodicities using the $Z^2_n$ test (Protheroe 1987). As the number of harmonics *n* that are summed together is increased, this test becomes more sensitive to narrower pulses. Without a signal, the $Z^2_n$ statistic asymptotically follows a $\chi^2$ distribution with 2*n* degrees of freedom. This property and the number of independent periods tested (eleven) were used to set the confidence levels.

A period search was performed using 1 d steps in period between 1 and 2000 d. The resulting periodogram (*n* = 5) with the 95% (200 – 2000 d) confidence level is shown in Fig. 2. There is a lone peak at ~445 d. This peak is still one of the largest with *n* = 2 (sinusoidal), or *n* = 10 (very narrow pulses). Given these possible period detections of a period we thought it worthwhile to search other data sets.

Since 1999, the PCA instrument (2 – 30 keV) on RXTE has been scanning the Galactic bulge once every 2 weeks. The relevant GX 1+4 data span MJD 51214 – 53189. We selected data with low (< 10 cts/s/5PCU) backgrounds. These data were mostly in the latter part of the mission i.e. after MJD 51847. The LSP analysis of the data is shown in Fig. 3**.** Because of the short on-source times the fluxes show a large amount of scatter caused by the high amplitude pulsations. This non-Poissonian scatter causes large powers. The limit shown is the 95% confidence white noise level, which is determined in the same way as for the ASM data. Because of the large scatter, and apparent large-scale variations red noise is a major factor. The significance of the peaks must thus be treated with caution.

The BATSE instrument, aboard the Compton Gamma-Ray Observatory, consisted of a set of 8 scintillation detectors designed to detect gamma-ray bursts. However, using the Earth occultation technique (Harmon et al. 2002) the ~ 20 to 100 keV fluxes of steady and transient sources could be determined. We combined the individual occultation steps into daily, 20 – 70 keV, flux averages for GX 1+4 for the interval

MJD 48407 – 51689. This interval partially overlaps the RXTE/ASM data. Fig. 4 shows the resultant periodogram of the BATSE data. The highest peaks are at 423, 540, and 758 d periods. The 758 d period can be discounted since it also appears in the window function. Again the possible presence of red-noise means that the apparent significance of these peaks against white noise limits must be used as a guide only.

*3.4 Discussion*

The lack of any significant detection in the 10 d averaged and flare-filtered RXTE/ASM data implies that there is no smooth flux modulation with > ~16% p-p amplitude in the range 1.5 – 12.0 keV at a ~304 d period. Note that the 304 d period reported by Pereira et al ((1999) was based on the periodic modulation of the pulsar frequency and is not directly tested by the analysis presented here. But we limit any flux modulation linked with the proposed 304 d period. In addition an upper limit (95% confidence) of 22% p-p amplitude over the range 20-2000 d can be set for any sinusoidal signal at 1.5 – 12.0 keV energies. However, when flares are included a period near 440 d with 25% p-p amplitude is tentatively detected. A search based upon the $Z^2_n$ statistic and flare times shows a similar period (~460 d). Similar period signals are also found in the other two data sets (PCA Galactic Centre scans, and BATSE). These results can be explained by semi-regular flaring with a quasi period of 420 to 460 d. The interval between flares would not be strictly periodic, and/or flares are skipped by the source. By analogy with Be transients this would imply that GX1+4 has a highly eccentric orbit (<10 day periastron passage on a ~300 d period) and that the accretion flow in not steady.

**4. Sco X-2 (GX 349+2)**

*4.1 Introduction*

Sco X-2 is a bright, flaring, neutron star LMXB and Z-source with an orbital period of 22.3 h. Other Z-sources are Sco X-1, Cyg X-2, GX 5-1, GX 17+2, and GX 340+0. A Z-track is traced out in the X-ray colour-colour diagram on timescales from seconds to days. To a greater or lesser extent all Z-sources show long-term shifts in position on colour-colour, and hardness-intensity diagrams. The best studied such source in this regard is Cyg X-2 which exhibits periodic flux variations (Paul et al. 2000) at ~40 d and ~70 d as the whole Z track moves. Sco X-2 also shows Z-track shifts (Kuulkers & van der Klis 1995) with the high state flux being 30% brighter than the low state flux. Our main objective was to ascertain if these variations are periodic.

*4.2 Observations, Analysis and Results*

About 6.3 years (MJD 50088 – 52390) of RXTE/ASM data were used for these analyses. To avoid any trends that would be introduced by detector gain changes we selected only SSC2 data, and data from SSC1 with a constant gain. To mitigate against flaring we used 10 d and 20 d averages. Moreover, we used only the lowest (A) energy band as the flaring was weakest in this band. Thus our period search using LSP analysis (Fig. 5) was for broad smooth variations in the overall low energy flux. No significant (95% and 99 % confidence level) peaks were found using either the 10 or 20 d averaged data. Note that the powers in Fig. 5 appear much lower than the detection level because this level accounts for the number of independent frequencies in the periodogram. We established an upper limit of ~ 15% p-p amplitude for sinusoidal modulations in the 30 – 100 d range. Also, there was no evidence (broad hump) in the periodogram for a preferred timescale.

*4.3 Discussion*

The upper limit of ~15% is less than the observed 30% full-amplitude changes in intensity as the Z-track moves in the colour-colour diagram. It appears that these changes for Sco X-2 are not periodic. Therefore a steady precessing accretion disk is unlikely to be producing the observed changes in overall flux of Sco X-2. This is consistent with the modelling of Ogilvie & Dubus (2001): a precessing disk would not be expected in Sco X-2, given its short orbital period. Instead, Sco X-2 probably lies in the so-called "indeterminate instability zone". In this case, aperiodic changes in overall flux might be explained as resulting from changes in mass accretion rate. A comprehensive search for the presence of aperiodic or quasi-periodic variations is required to test this possibility.

**5. GX 339-4**

*5.1 Introduction*

GX 339-4 is a firm black hole candidate because it exhibits all the canonical states associated with black hole systems that have reliably measured high mass functions, such as Cyg X-1 and GRO 1655-40. The source demonstrates highly irregular variability at all wavelengths, making it difficult to definitively determine the orbital period as any periodic phenomena tend to be easily obscured. A number of values for the orbit have been reported, ranging from 0.5 – 1.7 d. The source exhibits aperiodic and quasi-periodic variability on time-scales spanning milliseconds to years, with a ~190 – 240 d super-orbital period reported (Kong et al. 2002). This long-term variability is considered a characteristic time-scale rather than a strict clocking phenomenon, and is believed it to be the result of the precession of a radiatively warped accretion disk.

*5.2 Observations, Data Analysis, Results*

Here, we investigate both the reported orbital and super-orbital periods. Our investigation of the reported super-orbital period extends beyond the Off State (OS) and the Low State (LS), to include the High State (HS) and the Very High State (VHS). The RXTE/ASM light-curve, for data spanning MJD 50088 – 53047, shows the full range of states exhibited by GX 339-4 (Fig. 6). The identification of the various states (Kong et. al 2002) is shown. The two outbursts relate to the HS: MJD 50800 – 51200, and the VHS: MJD ~52370 – 52740. The LS is always transitioned as GX 339-4 enters or leaves a HS or a VHS. We used the LS data from MJD 50088 – 50800 during which time the ASM flux was ~2 cts/s. In the OS, the ASM count rate dropped below the 3σ detection level (mean flux ~0.1 count s$^{-1}$).

The ASM data for GX 339-4 were treated in a similar manner to those for Sco X-2, with data from the SSC3 data being rejected due to the degradation of the instrument over time (A. M. Levine 2005, private communication) and linear trends being removed to account for the calibration drifts.

Previously, Kong et al. (2002) and Zdziarski et al. (2004) had discarded data from the HS and VHS for their analyses, due to the power spectra being dominated by the large scale flux level changes as the outbursts associated with each state progressed. In order to make these data useable, we attempted to remove the affect of the outbursts (states) in the light curve by fitting and subtracting cubic polynomials from the HS and VHS sections of data. To effectively fit linear and polynomial trends to the data, the light curve sections were initially re-binned into 1 d averages.

Power spectra for each of the light curve sections were generated using the LSP technique. Our results confirm that the OS does indeed represent non-detection, as the resulting power spectrum clearly resembles that of noise-dominated data at high frequencies. In addition there is a complete absence of any power around the ~190 – 240 d super-orbital period identified by Kong et al. (2002). The power spectrum of the LS (Fig. 7, lower panel) shows a broad double peak of highly significant power (well above the 99.9% white noise significance level) at 200 ± 70 d FWHM consistent with the results of both Kong et al. (2002) and Zdziarski et al. (2004). The power spectrum of the trend-removed HS (Fig. 7, middle panel) shows a highly significant (against white noise) double peak at 170 ± 70 d FWHM. In contrast, the power spectrum of the trend-removed VHS (Fig. 7, upper panel) shows a single broad peak at 150 ± 40 d FWHM. Although the peaks are highly significant when tested against a white noise process, the presence of red-noise due to source flaring or changes in the variance of the data will change the power spectrum statistics and hence reduce the significance of the detections. Simulations of Kong et al. 2002 show that even when red-noise is taken into account the super-orbital period remains significant.

Initially thinking that the short duration of the HS and VHS light curve sections (~400 d) had shifted the value of the super-orbital period, we performed simulations to determine whether LSP analysis could detect a real modulation of ~200 d in these sections. Simulated HS and VHS light curves with the count rates replaced with a sine wave of period 210 d were analysed. While the period of the modulation as determined from the FWHM of the broad peak did seem to be shifted slightly, the LSP power spectra showed that a broad peak at ~200 – 210 d would still appear. Hence the short duration of the light curve sections does not significantly affect the location of the peak. While the width of the peaks could indeed indicate that the super-orbital variability is not strictly periodic, simulations have shown that the LSP analysis technique cannot constrain a periodic modulation of ~190 – 240 d to any greater certainty in data sets of this length (~400 - 700 d, i.e. only ~1 - 3 cycles).

Comparison of the LSP results from the LS, HS and VHS would seem to suggest a possible evolution of the super-orbital period correlated with the changes in state. The minor peak at ~ 140 d in the LS increases in significance with the transition to the HS, before merging with the ~200 d peak to form a single peak at ~ 150 d in the VHS. It has been assumed that the cubic polynomial trends account for the intensity changes due to the change of state, and that any red-noise (if present) is on time scales shorter than the state change.

Previous analyses have detected a highly significant spurious peak at ~1 d in the LSP power spectrum of GX 339-4 (e.g. Bennloch 2004; Zdziarski et al. 2004). This peak has recently been attributed to the spectral leakage of low-frequency power present in the light curve (see Farrell et al. 2005 for a full discussion). In an attempt to determine whether a real modulation with a 0.5 – 1.7 d orbital period is present in the ASM data, the light curve sections were also re-binned into 0.3 d averages so as to shift any spectral leakage signature away from 1 d.

LSP analysis of the light curve re-binned into 0.3 d averages showed that the peaks around ~1 d were indeed shifted to 0.3 d. Re-binning the data changed the effective sampling rate from ~1 d to 0.3 d, thus removing any spectral leakage signature from around the proposed orbital period. The resulting power spectra of the LS, HS and VHS light curve sections showed no evidence of any period greater than the 50% white noise significance level between 0.5 – 1.7 d (Fig. 8). Simulations were then performed to determine whether or not a period between 0.5 – 1.7 d could be detected by LSP analysis in a light curve re-binned into 0.3 d averages. LSP power spectra of the LS, HS and VHS light curves with the count rates replaced by sine waves with periods of 0.5, 0.9, 1.3 and 1.7 d showed a highly significant peak at each of the respective periods. At these high frequencies red-noise is not an issue and so the significance levels calculated using white noise, as shown in our figures, are accurate. Thus it can be concluded that if the orbital period of the binary does lie within this range, it is not present in the RXTE ASM light curve.

*5.3 Discussion*

There may be a threshold luminosity somewhere between the Off and Low States where the irradiation of the disk "switches on" the super-orbital period. We would thus expect the super-orbital period to be present during the High and Very High States. It is possible that magnetic activity on the surface of the donor star could lead to significant variations in the accretion rate, thus altering the disk size and leading to variations in the precession period (J. R. Murray 2005, private communication). Zdziarski et al. (2004) proposed that entry into the first soft (High) state altered the accretion flow so significantly that the precession pattern changed, in order to account for their non-detection of the super-orbital period in either the entire ASM data set or the soft outbursts. This theory could account for the possible evolution of the period with state changes that we have observed, although further work is required before a firm conclusion can be drawn.

**Acknowledgements**

We thank Dr. Colleen Wilson at NTSSC/NASA-MSFC for extracting and processing the BATSE Earth Occultation data on GX1+4. Craig Markwardt of GSFC generously gave us permission to use RXTE/PCA Galactic Centre scan data. This research has made use of data obtained through the High Energy Astrophysics Science Archive Research Centre Online Service, provided by the NASA/Goddard Space Flight Centre.

**Figure captions:**

**Fig. 1** – Lomb-Scargle periodogram of the 1 d binned, full energy, RXTE/ASM data for GX1+4. Flaring was not excluded. The horizontal line shows the 95% confidence level for white noise. The peak at ~440 d does not exceed the 99% confidence level.

**Fig. 2**: The $Z^2_5$ test period search for RXTE/ASM data for GX 1+4.

**Fig. 3**: Periodogram of the RXTE/PCA Galactic Centre scan data for GX 1+4. The high powers and the apparent high significance of many peaks are attributable to the non-Poisson nature of the data.

**Fig. 4**: Lomb-Scargle periodogram of the BATSE ~20 – 70 keV data derived from Earth occultation's, for GX 1+4. The data have a non-Poisson distribution and there is red noise present.

**Fig. 5**: Lomb-Scargle periodogram of 10 d binned, RXTE/ASM (SSC2 only) data for Sco X-2, for periods in the range 20 to 100 d

**Fig. 6**: Daily averaged RXTE/ASM light curve of GX 339-4 for MJD 50088 – MJD 53047 showing the various states. The linear and cubic polynomial trends that were subtracted from the LS, HS and VHS light curve sections prior to LSP analysis are shown.

**Fig. 7:** Lomb-Scargle periodogram of the Low State (lower panel), the High State (middle panel), and the Very High State (upper panel) for GX 339-4. The highest significance periods are 200 ± 70 d FWHM, 170 ± 70 d FWHM, and 150 ± 40 d FWHM respectively. The dashed lines indicate the 99% white-noise significance levels.

**Fig. 8:** Lomb-Scargle periodogram of the individual dwell (black) and 0.3 d re-binned (grey) light curve sections for the Low State (lower panels), High State (middle panels), and Very High State (upper panels) for GX 339-4, showing the removal of the ~1 d peaks by re-binning. The dashed lines indicate the 99% white-noise significance levels.

**Figures**

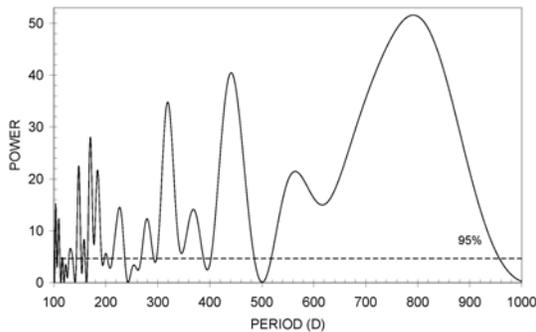
Fig. 1

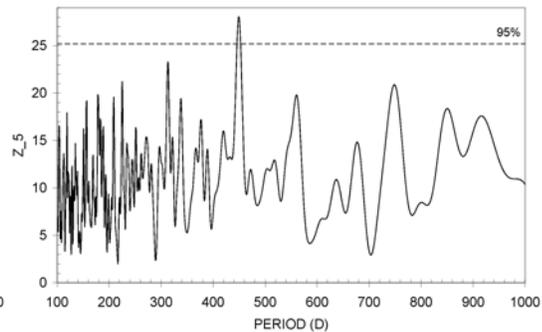
Fig. 2

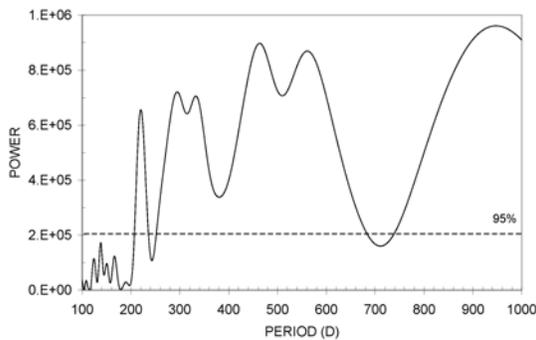
Fig. 3

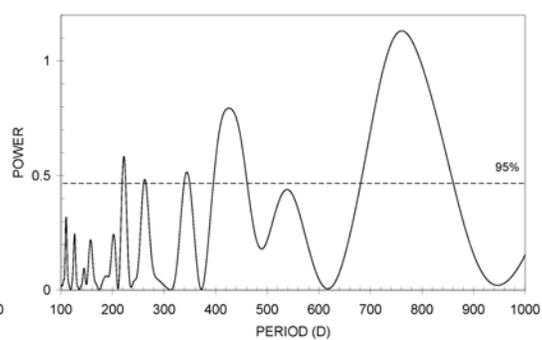
Fig. 4

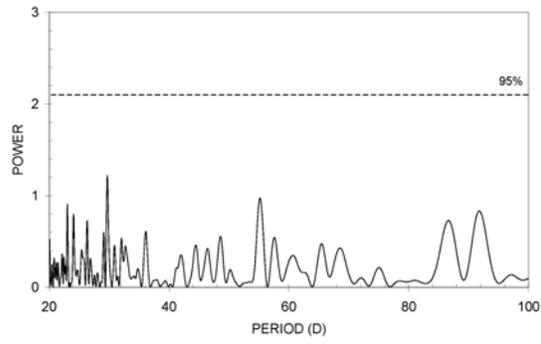
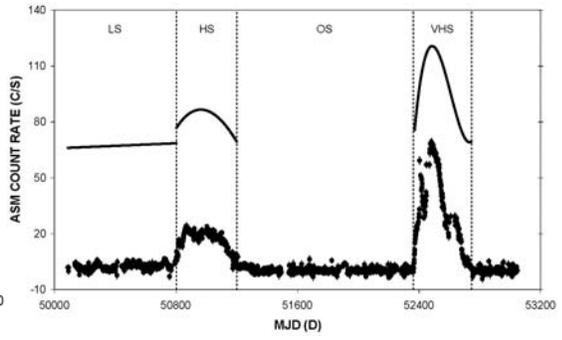

Fig. 5                                    Fig. 6

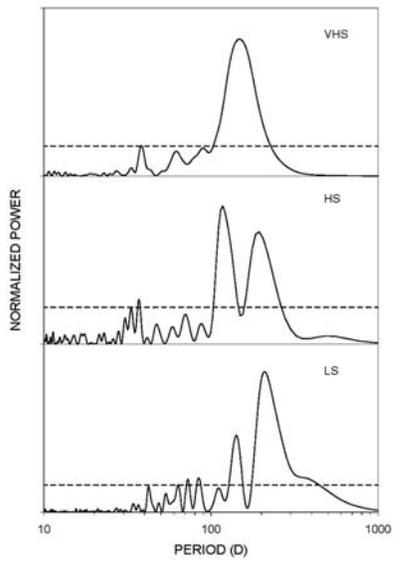
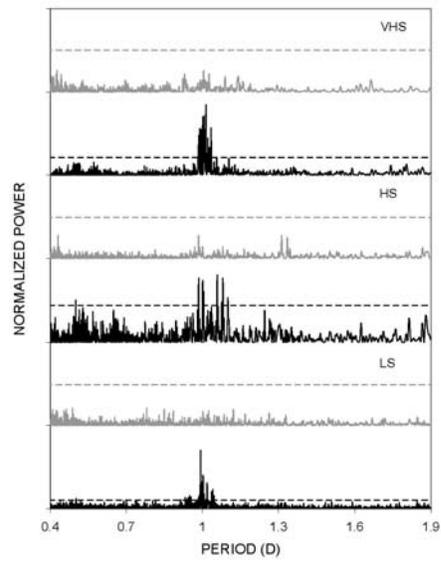

Fig. 7                                    Fig. 8